# Balancing Power, Efficiency, and Constancy under Broken Time-Reversal Symmetry

Ousi Pan[1], Zhiqiang Fan[1], Shunjie Zhang[1], Liwei Chen[2], Jincan Chen[1], Shanhe Su[1+]

[1]Department of Physics, Xiamen University, Xiamen 361005, People's Republic of China

[2]School of Mechanic and Electronic Engineering, Sanming University, Sanming 365000, China

**Abstract:** We derive general trade-off relations among the power, efficiency, and constancy for two-terminal thermoelectric systems in the linear response regime. Constancy, which quantifies the steadiness of the heat engine, is measured by its fluctuations. The bounds of the efficiency, power and fluctuations are valid even when time-reversal symmetry is broken, revealing how such a symmetry breaking alters the fundamental constraints on steady-state energy conversion. Our results extend and refine previously established universal trade-offs, offering deeper insight into the performance limits in nonequilibrium thermodynamics. Guided by this bound, heat engines with broken time-reversal symmetry can be operated at near-Carnot efficiency while maintaining finite power output and fluctuations, enabling them to outperform their traditional counterparts.

---

+ sushanhe@xmu.edu.cn

## I. Introduction

The Carnot efficiency $\eta_C$ sets the fundamental upper bound for the efficiency of the heat engines operating between two thermal reservoirs. However, the bound is unattainable for practical heat engines, as it demands an infinitely slow operation that yields vanishing power output. This limitation has therefore stimulated extensive research on the efficiency at maximum power [1]. In particular, the optimal efficiency and power have been explored in thermoelectric systems [2-6]. Some bounds linking the power and efficiency have been established [7-9], but the performance of an engine also depends critically on stability, characterized by the constancy of the power output. Initially derived from random processes within the linear response regime, the trade-offs among the power, efficiency, and constancy were subsequently extended to encompass more general and interacting systems [10]. These developments have deepened our understanding of the fundamental constraints that govern energy conversion far from equilibrium [11, 12].

The fundamental trade-offs among the power, efficiency, and constancy in nonequilibrium thermodynamic systems are elegantly captured by the thermodynamic uncertainty relations (TURs) [13], which bound current fluctuations in terms of entropy production. Specifically, the TUR can be expressed as [13-15]

$$Q_I = \frac{S_I \sigma}{I^2 k_B} \geq 2, \tag{1}$$

or

$$Q_J = \frac{S_J \sigma}{J^2 k_B} \geq 2, \tag{2}$$

where $I$ and $J$ denote the charge and heat currents of thermodynamic systems.

Their fluctuations, denoted as $S_I$ and $S_J$, characterize the system's constancy or stability. The entropy production rate is given by $\sigma$, and $k_B$ represents the Boltzmann constant. Originally formulated in the context of biomolecular processes, the TUR has since been extended to a broad class of thermodynamic systems, including finite-time thermodynamic processes [16, 17], Brownian motion [18], Langevin dynamics [19], and stochastic thermodynamics [20]. Its theoretical foundation has been rigorously established through the large deviation theory [14, 17, 21], and its predictions have been validated by experimental results [16, 22]. Due to its generality, the TUR has been connected to universal bounds on fluctuations [23-25] and to the fundamental limits governing the power, efficiency, and precision in thermodynamic machines [11, 12, 26-28]. However, deviations from the standard TUR form emerge in various scenarios, such as periodically driven systems [29-31] or overdamped Langevin dynamics with time-reversal-odd observables [32]. In particular, when time-reversal symmetry is broken, the generalized forms of the TUR have been proposed [15, 33-35]. The breakdown of time-reversal symmetry can dramatically alter the thermodynamic landscape, leading to novel bounds on fluctuations and efficiencies, especially in mesoscopic thermoelectric systems [36, 37].

Given that the form of the TUR varies across different physical systems, the associated bounds on the power, efficiency, and constancy are likewise system-dependent. In this work, we derive a tighter trade-off bound among these quantities in the linear response regime for systems with broken time-reversal symmetry, based on the generalized TUR proposed in Ref. [15]. Our analysis reveals how symmetry breaking can relax conventional constraints, enabling enhanced performance in thermodynamic machines.

To explore how time-reversal symmetry breaking reshapes thermodynamic constraints, we derive a tighter bound linking the power, efficiency, and constancy in the linear response regime, building on a generalized form of the TUR [15]. We begin by introducing the physical model and outlining the theoretical framework underpinning our analysis. We then evaluate the new bound under representative parameter settings and compare it with the established Pietzonka and Seifert's bound [11], highlighting the improvements enabled by symmetry breaking. Furthermore, we present an explicit model that breaks the time-reversal symmetry. Finally, we discuss the broader significance of our findings and their potential implications for the design of high-performance energy conversion systems.

**II. Model and Theory**

We consider a two-terminal setup with broken time-reversal symmetry induced by an external magnetic field $\vec{B}$, as illustrated in Fig. 1. The central region consists of a conductor coupled to two reservoirs. The left and right reservoirs are maintained at temperatures $T_L$ and $T_R$, respectively, with the corresponding voltage $V_L$ and $V_R$. The temperature difference $\Delta T \equiv T_L - T_R > 0$ and the voltage difference $\Delta V \equiv V_L - V_R$ are both small. Within the framework of linear response theory, the charge current $I$ and the heat current $J_h$ flowing out of the left reservoir are related to the thermodynamic forces via [38, 39]

$$\begin{pmatrix} I \\ J_h \end{pmatrix} = \begin{pmatrix} L_{11} & L_{12} \\ L_{21} & L_{22} \end{pmatrix} \begin{pmatrix} X_\mu \\ X_T \end{pmatrix},$$

where $L_{ij}$ $(i, j = 1, 2)$ are the Onsager coefficients and $X_\alpha$ $(\alpha = \mu, T)$ are the thermodynamic forces. Specifically, $X_\mu = \frac{\Delta V}{T_R}$ and $X_T = \frac{\Delta T}{T_R^2}$ in our regime. $L_{ij}$ satisfies the well-known Onsager-Casimir reciprocity relation $L_{ij}(\vec{B}) = L_{ji}(-\vec{B})$. This implies $L_{12} = L_{21}$ when time-reversal symmetry is preserved (i.e., $\vec{B} = 0$), and $L_{12} \neq L_{21}$ when time-reversal symmetry is broken (i.e., $\vec{B} \neq 0$). Under this framework, the entropy production rate can be expressed in terms of thermodynamic forces and the Onsager coefficients as

$$\sigma = IX_\mu + J_h X_T = L_{11} X_\mu^2 + L_{22} X_T^2 + (L_{12} + L_{21}) X_\mu X_T. \tag{3}$$

The second law of thermodynamics imposes a fundamental constraint on this expression, requiring that $\sigma \geq 0$ for all admissible forces [38]. To enforce this condition within the framework of Onsager's theory, it is useful to express the entropy production rate as

$$\sigma = \begin{pmatrix} X_\mu & X_T \end{pmatrix} \begin{pmatrix} L_{11} & \frac{L_{12} + L_{21}}{2} \\ \frac{L_{12} + L_{21}}{2} & L_{22} \end{pmatrix} \begin{pmatrix} X_\mu \\ X_T \end{pmatrix} = \begin{pmatrix} X_\mu & X_T \end{pmatrix} \mathbf{L}_+ \begin{pmatrix} X_\mu \\ X_T \end{pmatrix}.$$

The non-negativity of $\sigma$ for all values of $X_\mu$ and $X_T$ implies that the matrix $\mathbf{L}_+$ must be positive semi-definite. For a real symmetric matrix, this condition is ensured by all eigenvalues being non-negative. Let $\upsilon$ denote an eigenvalue of $\mathbf{L}_+$. Then it satisfies the characteristic equation

$$\upsilon^2 - (L_{11} + L_{22}) \upsilon + L_{11} L_{22} - (L_{12} + L_{21})^2 / 4 = 0.$$

The discriminant of this quadratic equation equals $(L_{11} - L_{22})^2 + (L_{12} + L_{21})^2$, which is always non-negative. The eigenvalue $\upsilon$ will be non-negative if and only if

$L_{11} + L_{22} \geq 0$ and $L_{11}L_{22} - (L_{12} + L_{21})^2/4 \geq 0$ hold. Hence, it can be derived that

$$L_{11} \geq 0 \text{ and } L_{22} \geq \frac{(L_{12} + L_{21})^2}{4L_{11}} \geq 0. \tag{4}$$

We consider the case where the system is operated as a heat engine under the conditions $\Delta T > 0$ and $\Delta V < 0$. The temperature difference $\Delta T$ drives the diffusion of charge carriers, generating a heat current $J_h$ flowing out of the hot reservoir $L$ and a simultaneous electric current $I$ that flows against the biased voltage $\Delta V$. This phenomenon enables the conversion of thermal energy into useful electrical power, which is given by the product of the voltage and the electric current. Another key performance metric for the heat engine is the efficiency, defined as the ratio of the useful work output to the heat input. Therefore, the power output and efficiency of the devices are given by

$$P = -\Delta V I, \tag{5}$$

and

$$\eta = \frac{P}{J_h}, \tag{6}$$

respectively, where the negative sign in Eq. (5) results from the opposing direction of the biased voltage and the electric current. The same system can also be operated as a refrigerator, in which electrical power is consumed to pump heat from the cold to the hot reservoir, or as a heat dissipator, where electrical power is absorbed and heat flows from the hot reservoir to the cold one, depending on the value of $\Delta V$ [38]. Conventionally, this voltage difference $\Delta V$ is given by the fundamental relation $\Delta V = \Delta \mu / (-e)$, where $\Delta \mu = \mu_L - \mu_R$ denotes the electrochemical potential difference and $e$ is the elementary charge. To achieve a positive power output, the charge current

$I$ must be positive, which imposes an additional constraint on the applied voltage, resulting in

$$-\frac{L_{12}\Delta T}{L_{11}T_R} < \Delta V < 0. \tag{7}$$

Therefore, the coefficient $L_{12}$ must be positive for the system to function as a heat engine. For values of $\Delta V$ outside this range, the system may instead behave as a refrigerator or a heat dissipator, as described previously. Using Eqs. (5) and (6) to rewrite the entropy production rate, we obtain

$$\sigma = I\frac{\Delta V}{T_R} + J_h\frac{\Delta T}{T_R^2} = \frac{P}{T_R}\left(\frac{\eta_C}{\eta} - 1\right), \tag{8}$$

where $\eta_C = \Delta T / T_R$ is the Carnot efficiency [36-38].

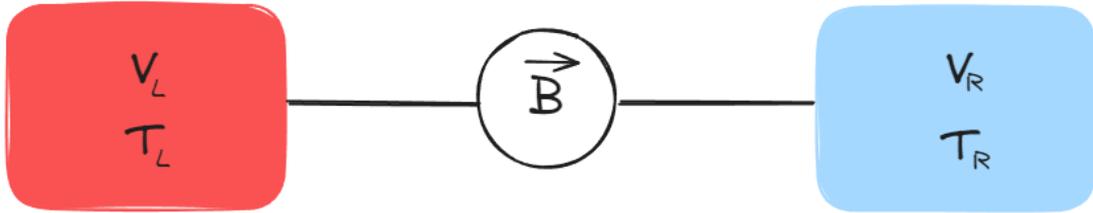

FIG. 1. The sketch of a two-terminal system. The central conductor, subjected to a magnetic field, is connected to two reservoirs characterized by a temperature difference $\Delta T$ and a voltage difference $\Delta V$.

In contrast to the conventional TUR in Eqs. (1) and (2), the TURs for two-terminal systems with broken time-reversal symmetry can be expressed as [15]

$$Q_I \geq 2\left[1 + \frac{1}{2}\frac{L_{21} - L_{12}}{L_{12} + L_{11}\left(\frac{\Delta V T_R}{\Delta T}\right)}\right]^2, \tag{9}$$

and

$$Q_J \geq 2\left[1 + \frac{1}{2}\frac{L_{12} - L_{21}}{L_{21} + L_{22}\left(\frac{\Delta T}{\Delta V T_R}\right)}\right]^2. \tag{10}$$

Note that we have used $S_I = 2k_B L_{11}$ and $S_J = 2k_B L_{22}$ based on the fluctuation-dissipation relations in the linear response regime at equilibrium [15, 28]. The fluctuations $S_I$ and $S_J$ reflect the constancy or stability of the system. The greater the fluctuation amplitude is, the more unstable the system becomes. By employing Eqs. (5), (6) and (8), the quantities $Q_I$ and $Q_J$, originally defined in Eqs. (1) and (2), can be rewritten in terms of the entropy production rate, efficiency, and power as follows:

$$Q_I = \frac{S_I (\Delta V)^2 (\eta_C - \eta)}{P \eta k_B T_R}, \tag{11}$$

and

$$Q_J = \frac{S_J (\eta_C - \eta)\eta}{P k_B T_R}. \tag{12}$$

A combination of Eqs. (9) and (11) yields

$$P \frac{\eta}{\eta_C - \eta}\frac{k_B T_R}{S_P} \leq C_1, \tag{13}$$

where $C_1 = \{1 + (L_{21} - L_{12})/[2(L_{12} + L_{11}\Delta V T_R / \Delta T)]\}^{-2}/2$. The term $S_P = S_I (\Delta V)^2$ in the denominator on the left of the inequality represents the power fluctuations in the thermoelectric system [24, 28].

Likewise, combining Eqs. (10) and (12) yields another key inequality

$$\frac{P k_B T_R}{(\eta_C - \eta)\eta S_J} \leq C_2, \tag{14}$$

where $C_2 = \{1 + (L_{12} - L_{21})/[2(L_{21} + L_{22}\Delta T / (\Delta V T_R))]\}^{-2}/2$. In the presence of time-reversal symmetry, i.e., when $L_{12} = L_{21}$, Eq. (14) reduces to the trade-off relation mentioned in Ref. [11]. However, our result is extended to the systems with broken time-reversal symmetry.

**III. Results and Discussion**

The performance of a heat engine cannot be assessed solely by the power or efficiency alone without considering the constancy. Consequently, fluctuations are also critical in our analysis. To emphasize the difference between our findings and those for systems with time-reversal symmetry, we will focus on the bound given in Eq. (13) and compare it with the Pietzonka-Seifert bound, which satisfies $k_B T_R P \eta / \left[ (\eta_C - \eta) S_P \right] \leq 1/2$ [11]. A similar procedure can be applied to Eq. (14), yielding an analogous result.

Since the form of Eq. (13) is similar to that of the Pietzonka-Seifert bound, we only rewrite $C_1$ as

$$C_1 = \frac{1}{2} \left[ 1 + \frac{l_{21} - 1}{2(1 + l_{11}\alpha)} \right]^{-2}, \tag{15}$$

where the dimensionless parameters $l_{11} = k_B T_R L_{11} / (e L_{12})$, $l_{21} = L_{21} / L_{12}$, and $\alpha = e\Delta V / (k_B \Delta T)$. To simplify the notation, we define $l_{11} = f L_{11} / L_{12}$, where $f = k_B T_R / e$. From one perspective, parameter $l_{21}$ determines whether the bound reduces to the Pietzonka-Seifert bound (for a system without broken time-reversal symmetry, $l_{21} = 1$); from another, $l_{11}\alpha$ in the denominator also has an effect on the bound. For example, when $\alpha$ becomes infinite (i.e. $\Delta T = 0$), the bound again reduces to the Pietzonka-Seifert bound.

The Pietzonka-Seifert bound implies that it is impossible to achieve a finite power,

efficiency close to the Carnot limit, and constancy (low power fluctuations) simultaneously [11]. Indeed, achieving Carnot efficiency at finite power generally demands extreme conditions—such as an unlimited number of nonreciprocal energy conversion units [40], an infinite number of energy recycling processes [41], or infinitesimal relaxation times [42, 43]. Yet, according to our bound, for parameters satisfying $l_{21} \approx 1 - 2(1 + l_{11}\alpha)$, a stable heat engine operating close to the Carnot efficiency while delivering nonzero power output is theoretically attainable under broken time-reversal symmetry. Furthermore, the careful tuning of parameters enables a heat engine under broken time-reversal symmetry to outperform a traditional one. Thus, our analysis goes beyond the work of Ref. [36] by incorporating fluctuations, extends the results of Ref. [11] to regimes with broken time-reversal symmetry, and further generalizes the theoretical framework of Ref. [15] to demonstrate that the performance of heat engines can be enhanced under broken time-reversal symmetry.

To illustrate our results, we model an ideal heat engine subject to the constraints specified in Eqs. (4), (7), (13), and (14). The following parameter choices are adopted. The parameter $l_{11}$ is set to four different values: $0.05f$, $0.10f$, $0.15f$, $0.20f$. To prevent $C_1$ from diverging in our calculations, we take $l_{21} = -2l_{11}\alpha - 1.1$. To satisfy Eq. (4), we set $L_{22} = 0.2525(L_{12} + L_{21})^2 / L_{11}$ $\left[ k_B^2 T_R^3 / h \right]$, where $h$ is Planck's constant. Within the linear response regime, a small temperature difference $\Delta T = 0.1 T_R$ is applied, ensuring $\Delta T$ is small compared to $T_R$. Additionally, we fix $L_{12} = 0.2$ $\left[ ek_B T_R^2 / h \right]$. The maximum power output, which is achieved under the condition $l_{11}\alpha = -1/2$, is given by [36]

$$P_{\max} = \frac{\eta_C}{4} \frac{L_{12}^2}{L_{11}} X_T. \tag{16}$$

Using the definition of the power fluctuations $S_P = 2k_B L_{11}(\Delta V)^2$, the optimal working condition $l_{11}\alpha = -1/2$, and Eq. (16), we derive the power fluctuations of the system at maximum power output as

$$S_{P,\text{MP}} = 2k_B T_R P_{\max}. \tag{17}$$

Fig. 2(a) shows that as $\alpha$ decreases (corresponding to an increase in the voltage difference $\Delta V$), the normalized efficiency $\eta/\eta_C$ rapidly rises to nearly unity and remains close to 1 across a wide range of $\alpha$ before eventually falling to zero. The normalized power $P/P_{\max}$ also exhibits a non-monotonic behavior, first increasing and then decreasing after reaching the top, as illustrated in Fig. 2(b). From Fig. 2(c), it is evident that the power fluctuations remain finite within the chosen parameter range. Fig. 2(d) displays the variation of $l_{21}$, which reflects the extent of broken time-reversal symmetry. The results in Fig. 2(e) confirm the validity of Eq. (13): the curves lying above the double black line indicate violation of the Pietzonka-Seifert bound, whereas our proposed bound is satisfied throughout. The correctness of Eq. (14) has also been verified (not shown). Finally, Fig. 2(f) demonstrates that efficiencies close to the Carnot-limit can be achieved while maintain finite power output, and finite power fluctuations as supported by Fig. 2(c).

IV. Aharonov-Bohm ring with broken time-reversal symmetry

Several concrete models have been proposed as potential means to break time-reversal symmetry [5, 15, 44-49]. To further illustrate our results, we consider a model

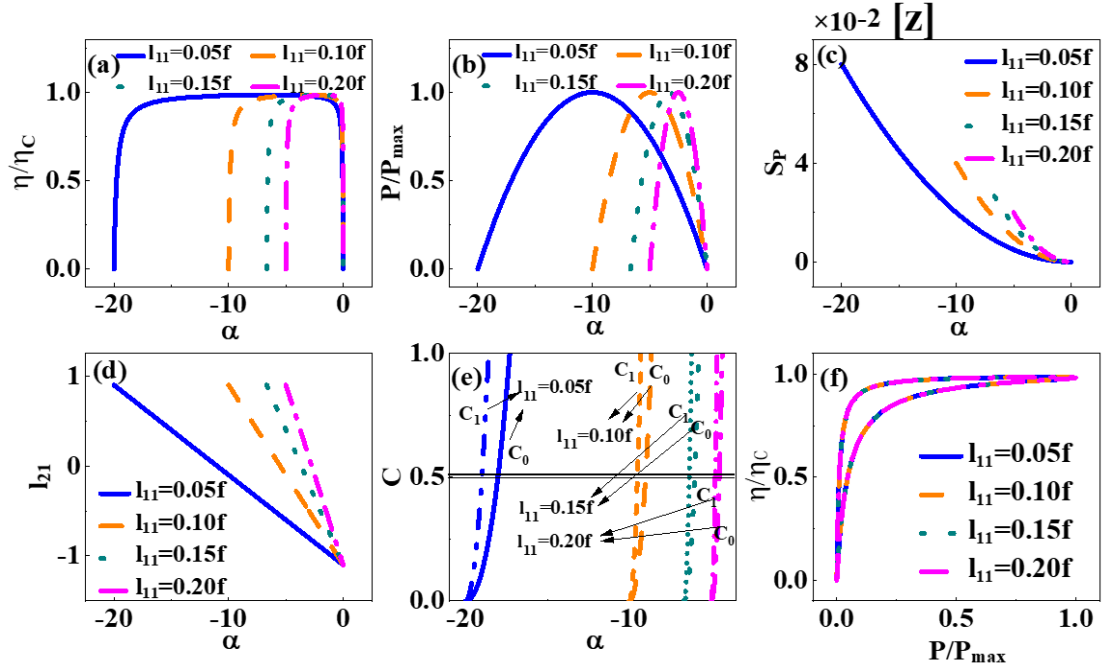

FIG. 2. (a) The normalized efficiency $\eta/\eta_C$, (b) the normalized power $P/P_{\max}$, (c) the power fluctuations $S_P$ in units of $z$ (where $z = k_B^3 T_R^3 / h$), and (d) the dimensionless ratio $l_{21}$ plotted as functions of the dimensionless ratio $\alpha$ under different parameter settings. The curves correspond to $l_{11} = 0.05f$ (solid blue), $l_{11} = 0.10f$ (dashed orange), $l_{11} = 0.15f$ (dotted teal), and $l_{11} = 0.20f$ (dash-dotted magenta). (e) The left-hand side (LHS, $C_0$) and the right-hand side (RHS, $C_1$) of Eq. (13) versus the dimensionless ratio $\alpha$. The double black line denotes the Pietzonka-Seifert bound. (f) The normalized efficiency $\eta/\eta_C$ plotted as a function of the normalized power $P/P_{\max}$ under different parameter settings. The curve representations are the same as those in (a)-(d).

similar to that in Refs. [5, 45, 46], as shown in Fig. 3(a). The system consists of an Aharonov-Bohm ring embedded between two electronic reservoirs. These reservoirs are biased at electrochemical potentials $\mu_L$ and $\mu_R$, and maintained at temperatures

$T_L$ and $T_R$. Situated on the ring's upper arm, a quantum dot (the purple circle in Fig. 3(a)) couples the electronic system to the phonon reservoir by mediating electron-phonon interactions. A magnetic flux $\Phi$ pierces the ring. The temperature and electrochemical potential difference of two electronic reservoirs can be expressed as

$$\begin{aligned} T_L - T_R &= \Delta T, \\ \mu_L - \mu_R &= e\Delta V. \end{aligned} \quad (18)$$

In the linear-response regime, $\Delta T$ is small compared to $T = (T_L + T_R)/2$. The temperature of the phonon reservoir, denoted by $T_P$, is set to be slightly different from the base temperature $T$ and is defined by the relation

$$T_P = T + \Delta T_P. \quad (19)$$

The quantum dot comprises a single localized electronic level at energy $\varepsilon_0$, with vibrational modes characterized by the frequency $\omega_0$. We neglect interactions among electrons, whereas the electron-phonon coupling strength is denoted by $\gamma$. Then, the Hamiltonian of the dot is

$$H_d = \left[\varepsilon_0 + \gamma\left(b + b^\dagger\right)\right]c_0^\dagger c_0 + \omega_0\left(b^\dagger b + \frac{1}{2}\right), \quad (20)$$

where $c_0^\dagger (c_0)$ is the electronic creation (annihilation) operator on the localized level, and $b^\dagger (b)$ creates (annihilates) a phonon of frequency $\omega_0$ and we set $\hbar = 1$. The Hamiltonian for the leads is composed of three parts: the left lead, the right one and the direct coupling between them through the ring's lower arm. Its explicit form is

$$H_\text{lead} = \sum_L \varepsilon_L c_L^\dagger c_L + \sum_R \varepsilon_R c_R^\dagger c_R + \sum_{LR}\left(V_{LR} e^{i\Phi} c_L^\dagger c_R + \text{H.c.}\right), \quad (21)$$

where the operator $c_{L/R}^\dagger (c_{L/R})$ creates (annihilates) an electron of energy $\varepsilon_{L/R}$ in the left or right lead, and $V_{LR}$ denotes the tunneling matrix element for electrons hopping through the lower arm of the ring. The tunneling between the dot and the leads is described by the Hamiltonian

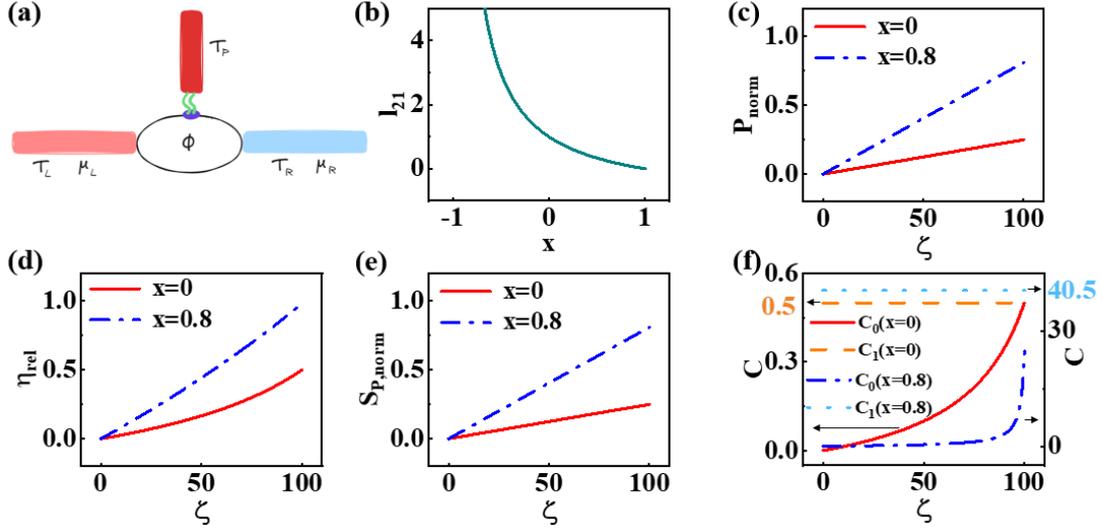

FIG. 3. (a) The sketch of a three-terminal system consisting of two electronic reservoirs at temperatures $T_L$ and $T_R$, and electrochemical potentials $\mu_L$ and $\mu_R$ together with a phonon reservoir at temperature $T_P$, interconnected by an Aharonov-Bohm ring. A quantum dot positioned on the upper arm of the ring enables electron-phonon coupling, allowing the phonon reservoir to influence the electronic system. (b) The parameter $l_{21}$ versus the parameter $x$. (c) The normalized power $P_{norm} = P_{max} / P_{max, global}$ versus the figure of merit $\zeta$. (d) The relative efficiency $\eta_{rel}$ versus the figure of merit $\zeta$. (e) The normalized power fluctuations $S_{P,norm}$ versus the figure of merit $\zeta$. In panels (c)-(e), the solid red and dash-dotted blue curves represent $x = 0$ and $x = 0.8$, respectively. (f) The left-hand side (LHS, $C_0$) and the right-hand side (RHS, $C_1$) of Eq. (38) versus the figure of merit $\zeta$. The solid red and dashed orange curves denote $C_0$ and $C_1$ at $x = 0$, respectively, while the dash-dotted blue and dotted light blue curves denote $C_0$ and $C_1$ at $x = 0.8$, respectively. For panels (c)-(f), we set $\zeta_{max} = 100$.

$$H_{\text{coup}} = \sum_L \left(V_L c_L^\dagger c_0 + \text{H.c.}\right) + \sum_R \left(V_R c_R^\dagger c_0 + \text{H.c.}\right), \tag{22}$$

where $V_L(V_R)$ describes the tunneling matrix element for electrons hopping between the dot and the left (right) lead. The total Hamiltonian consists of the above three components

$$H_{\text{total}} = H_d + H_{\text{lead}} + H_{\text{coup}}. \tag{23}$$

In the following, we introduce some parameters relevant to our model. The resonance width from the coupling between the quantum dot and the left (right) reservoir is given by $\Gamma_L(\Gamma_R)$, while energy-dependent coupling strength $\lambda(\omega)$ quantifies the direct coupling between the two leads. For the ring's lower arm, the electron transmission and reflection amplitudes are given by [45]

$$t_0^2(\omega) = \frac{4\lambda(\omega)}{\left[1+\lambda(\omega)\right]^2}, \quad r_0^2(\omega) = 1 - t_0^2(\omega). \tag{24}$$

The total resonance width is defined as [45]

$$\Gamma(\omega) = \frac{\Gamma_L(\omega) + \Gamma_R(\omega)}{1+\lambda(\omega)}. \tag{25}$$

The charge current $I$ and the electronic heat current $J_h^E$ flowing from the left reservoir, as well as the phonon-mediated heat current $J_h^P$ from the phonon reservoir, satisfy the following relation [5, 50]

$$\begin{pmatrix} I \\ J_h^E \\ J_h^P \end{pmatrix} = \begin{pmatrix} G(\Phi)T & K(\Phi)T & X^P(\Phi)T \\ K(-\Phi)T & K_2(\Phi)T & \tilde{X}^P(\Phi)T \\ X^P(-\Phi)T & \tilde{X}^P(-\Phi)T & C^P(\Phi)T \end{pmatrix} \begin{pmatrix} \Delta V/T \\ \Delta T/T^2 \\ \Delta T_P/T^2 \end{pmatrix}. \tag{26}$$

Here, $G$ is the electric conductance. $K$ is the coefficient characterizing the Seebeck effect. $K_2$ is the diagonal electronic thermal transport coefficient. $X^P$ and $C^P$ are coefficients with respect to the phonon-mediated analogues of $K$ and $K_2$. $\tilde{X}^P$ is the coefficient originating from the electron-vibrational interaction. Within the Keldysh Green functions formalism, the transport coefficients can be derived explicitly [51]. We adopt the specific forms presented in Refs. [45, 50].

To reduce the system to a standard two-terminal setup, we set $J_h^E = 0$ [5] by choosing

$$\Delta T = -\frac{T}{K_2(\Phi)}\left[K(-\Phi)\Delta V + \tilde{X}^P(\Phi)\frac{\Delta T_P}{T}\right]. \tag{27}$$

Eq. (26) can be rewritten as

$$\begin{pmatrix} I \\ J_h^P \end{pmatrix} = \begin{pmatrix} G(\Phi)T - \dfrac{K(\Phi)K(-\Phi)}{K_2(\Phi)}T & X^P(\Phi)T - \dfrac{K(\Phi)\tilde{X}^P(\Phi)}{K_2(\Phi)}T \\ X^P(-\Phi)T - \dfrac{K(-\Phi)\tilde{X}^P(-\Phi)}{K_2(\Phi)}T & C^P(\Phi)T - \dfrac{\tilde{X}^P(-\Phi)\tilde{X}^P(\Phi)}{K_2(\Phi)}T \end{pmatrix} \begin{pmatrix} \dfrac{\Delta V}{T} \\ \dfrac{\Delta T_P}{T^2} \end{pmatrix}. \tag{28}$$

Since $K_2$ is an even function of the magnetic flux [45] for sufficiently small $\Delta V$ and $\Delta T$, it follows that $L_{12}(\Phi) = L_{21}(-\Phi)$. Under the condition of zero electronic heat current $J_h^E = 0$, the Carnot efficiency of this heat engine is $\eta_C = \dfrac{\Delta T_P}{T}$ [52]. This result is consistent with our discussion in Sec. II.

We investigate the special case of symmetric coupling to the leads, i.e., $\Gamma_L = \Gamma_R$. Besides, we assume the wide-band approximation, so that $\Gamma_L(\omega) = \Gamma_L$, $\Gamma_R(\omega) = \Gamma_R$, $t_0(\omega) = t_0$, and $r_0(\omega) = r_0$. Under this condition, and following Ref. [5, 50], the coefficients $G$, $K$, $K_2$, and $C^P$ are even function of the magnetic flux $\Phi$, whereas $X^P$ is an odd function. Equation (28) can be reduced to the following simplified form [5, 50]:

$$\begin{pmatrix} I \\ J_h^P \end{pmatrix} = \begin{pmatrix} \dfrac{G}{1+\zeta}T & \dfrac{C^P K}{2K_2}(1+x)T \\ \dfrac{C^P K}{2K_2}(1-x)T & \dfrac{(C^P)^2}{4K_2}(\zeta_{\max} + x^2)T \end{pmatrix} \begin{pmatrix} \dfrac{\Delta V}{T} \\ \dfrac{\Delta T_P}{T^2} \end{pmatrix}, \tag{29}$$

where the argument $\Phi$ is suppressed for brevity. Here, $\zeta = K^2/(GK_2 - K^2)$ is the figure of merit for a conventional two-terminal thermoelectric heat engine, $\zeta_{max} = -1 + 4K_2/C^P$ denotes the theoretical upper bound of $\zeta$, and $x = t_0 \sin\Phi$. In such case, $l_{21} = L_{21}/L_{12} = (1-x)/(1+x)$. In Fig. 3(b), we plot $l_{21}$ as a function of $x$, where $x$ characterizes the variation of the magnetic flux $\Phi$. The dimensionless quantity $l_{21}$ exhibits a monotonic decrease with increasing $x$, and is exactly equal to 1 if and only if $x = 0$.

We next apply our conclusion from Sec. III to this model. The bound $C_1$ can be expressed as follows:

$$C_1 = \frac{1}{2}\left(\frac{1+x+\kappa}{1+\kappa}\right)^2, \tag{30}$$

where $\kappa = 2K_2 G\Delta V / [KC^P(1+\zeta)\eta_C]$. We focus on the scenario where the system operates at the maximum power. This condition is achieved when

$$\Delta V_{MP} = -\frac{(1+\zeta)(1+x)C^P K}{4GK_2}\eta_C. \tag{31}$$

Inserting Eq. (31) into Eq. (30), we obtain

$$C_1 = \frac{1}{2}\left(\frac{1+x}{1-x}\right)^2. \tag{32}$$

According to Eq. (16), the maximum power of this model is

$$P_{max} = K_2(\eta_C)^2 \frac{\zeta(1+x)^2}{(\zeta_{max}+1)^2}. \tag{33}$$

The global upper bound of $P_{max}$, denoted as $P_{max,\,global}$, is achieved exclusively at $\zeta = \zeta_{max}$ and $x = 1$, with the explicit form

$$P_{max,\,global} = \frac{4K_2(\eta_C)^2 \zeta_{max}}{(\zeta_{max}+1)^2}. \tag{34}$$

To simplify the subsequent analysis, we introduce the normalized power, defined as

$$P_{\text{norm}} = \frac{P_{\max}}{P_{\max,\text{global}}} = \frac{\zeta(1+x)^2}{4\zeta_{\max}}. \tag{35}$$

The power fluctuations are given by Eq. (17). Similar to the power output, we can also define the normalized power fluctuations as

$$S_{P,\text{norm}} = \frac{S_{P,\text{MP}}(P_{\max})}{S_{P,\text{MP}}(P_{\max} = P_{\max,\text{global}})} = \frac{\zeta(1+x)^2}{4\zeta_{\max}}. \tag{36}$$

The efficiency at maximum power is given by [5, 50]

$$\eta_{\text{MP}} = \frac{\eta_C}{2}\frac{L_{12}^2}{2L_{11}L_{22} - L_{12}L_{21}} = \frac{\eta_C}{4}\frac{(1+x)^2\zeta}{\zeta_{\max} + x^2 - \frac{1}{2}(1-x^2)\zeta}. \tag{37}$$

Combining Eqs. (13), (17), (33), and (37), it can be derived that

$$\frac{1}{2}\frac{\eta_{\text{rel}}}{1-\eta_{\text{rel}}} \leq C_1, \tag{38}$$

where the relative efficiency $\eta_{\text{rel}}$ is defined as $\eta_{\text{rel}} = \eta_{\text{MP}}/\eta_C$. Equation (38) is always satisfied, as the conditions $|x| \leq 1$ and $\zeta \leq \zeta_{\max}$ are inherently fulfilled by the physical constraints of the system.

In Fig. 3(c), we plot the normalized power $P_{\text{norm}}$ as a function of the figure of merit $\zeta$, which exhibits a monotonic increase with growing $\zeta$. A monotonic rise with $\zeta$ is also observed for the relative efficiency $\eta_{\text{rel}}$ in Fig. 3(d). In both panels, the dash-dotted blue curves $(x=0.8)$, corresponding to the scenario with broken time-reversal symmetry, lie above the solid red curves $(x=0)$, corresponding to the system with preserved time-reversal symmetry. Therefore, with broken time-reversal symmetry, it is possible to improve both the power output and the efficiency of the system. In Fig. 3(e), the normalized power fluctuations $S_{P,\text{norm}}$ likewise increase monotonically with $\zeta$. As indicated by Eq. (17), the power fluctuations remain finite as long as the power output is finite. For the parameter set $\zeta_{\max} = 100$, $\zeta = 100$, and $x = 0.8$, we achieve an efficiency $\eta = 98\%\eta_C$, approaching the Carnot limit. In this regime, Fig. 3(c) confirms a nonzero power output accompanied by finite power fluctuations, verifying that the near-Carnot efficiency, finite power output and finite fluctuations are mutually compatible in the system. We further validate our derived bounds in Fig. 3(f). For $x = 0$ (zero magnetic flux, $\Phi = 0$), our bound (the dashed orange curve) reduces exactly to the Pietzonka-Seifert bound, which lies above the solid red curve for all valid $\zeta$, confirming that the Pietzonka-Seifert bound is applicable for the scenario with preserved time-reversal symmetry under zero magnetic field. For $x = 0.8$, the dotted light blue curve lies above the dash-dotted blue curve, confirming the validity of our bounds for the broken time-reversal symmetry case. The dash-dotted blue curve, plotted against the right-hand y-axis, takes significantly larger values than the dashed orange Pietzonka-Seifert bound (plotted against the left-hand y-axis) for large $\zeta$. This result shows that the classical Pietzonka-Seifert bound is no longer valid for general scenarios with broken time-reversal symmetry, whereas our proposed bound remains well satisfied in both regimes with preserved and broken time-reversal symmetry.

As derived in Eq. (37), achieving the Carnot limit in the model requires an infinite $\zeta_{max}$, along with the conditions $\zeta = \zeta_{max}$ and $x = 1$. Yet this set of conditions simultaneously results in vanishing power output according to Eq. (33). While this model cannot reach the Carnot limit with nonzero power output at the same time, it enables stable operation at near-Carnot efficiency with finite power output and finite power fluctuations, outperforming conventional heat engines.

In addition, we note that enhancing the figure of merit is critical for boosting the performance of thermoelectric devices. However, the highest experimentally reported figure of merit is approximately 3, with a maximum documented value of 3.3 in Ref. [53], far below the range of values explored in our discussion. This further confirms that improving the figure of merit is a key pathway to substantially enhance the performance of practical thermoelectric heat engines.

## V. Conclusions

In summary, for two-terminal thermoelectric systems operating in the linear response regime, we have derived fundamental bounds that interrelate the power, efficiency, and constancy. The constancy is measured by the fluctuations in either the power output or the heat current. These bounds remain valid even when time-reversal symmetry is broken. Our results are based on the thermodynamic uncertainty relations established in previous relevant literature. In the presence of time-reversal symmetry, the bounds reduce to the well-known Pietzonka-Seifert bounds.

Phonon-assisted transport has been shown to enhance the efficiency of thermoelectric heat engines [47, 49]. Our findings further demonstrate that heat engines with broken time-reversal symmetry can achieve efficiencies close to the Carnot limit while simultaneously delivering finite power output and appreciable constancy (i.e., finite fluctuations). This combination of performance metrics exceeds the established capabilities of traditional heat engines. While our analysis is confined to the linear response regime, the general relationships among the power, efficiency and constancy in broader thermoelectric systems remain an open question. Although Ref. [54] has derived relations among the power, efficiency, and power fluctuations without broken time-reversal symmetry, the corresponding heat current fluctuations have not been incorporated. Consequently, a general TUR applicable to arbitrary currents is required to provide a better framework for the design of advanced heat engines. Furthermore, broken time-reversal symmetry has been experimentally observed [55], beyond the specific theoretical model validated in our work. Harnessing this principle to design superior thermoelectric heat engines remains a significant yet challenging goal for future work. Success in this endeavor would enable more effective conversion of waste heat from vehicles and factories into useful electricity.

**Acknowledgments**

This work has been supported by Natural Science Foundation of Fujian Province (2023J01006), National Natural Science Foundation of China (12364008 and 2023J01162), and Fundamental Research Fund for the Central Universities (20720240145) of China.